\documentclass[prl,final,twocolumn,showpacs]{revtex4}

\usepackage{graphicx}
\usepackage{dcolumn}

\begin{document}

\title{Origin of anomalously long interatomic distances in suspended gold chains}

\author{Sergio B. Legoas\footnote{Author to whom correspondence should be addressed.
FAX:+55-19-37885376.
Electronic address:slegoas@ifi.unicamp.br}$^{,1}$}
\author{Douglas S. Galv\~ao$^{1}$}
\author{Varlei Rodrigues$^{1,2}$}
\author{Daniel Ugarte$^{2}$}

\affiliation{$^{1}$Instituto de F\'{\i}sica "Gleb Wataghin", Universidade Estadual de
Campinas, C.P. 6165, 13083-970 Campinas SP, Brazil}
\affiliation{$^{2}$Laborat\'{o}rio Nacional de Luz S\'{\i}ncrotron, C.P. 6192, 13084-971 Campinas SP, Brazil}

\date{\today}

\begin{abstract}
The discovery of long bonds in gold atom chains has represented a challenge for physical interpretation. In fact, 
interatomic distances frequently attain 3.0-3.6 {\AA}
values and, distances as large as 5.0 {\AA} may be seldom observed. Here, we 
studied gold chains by transmission electron microscopy and performed theoretical calculations using 
cluster $ab$ $initio$ density functional formalism. We show that the insertion of two carbon atoms is required 
to account for the longest bonds, while distances above 3 {\AA} may be due to a mixture of clean and one C atom 
contaminated bonds. 
\end{abstract}

\clearpage

\maketitle

In the last years considerable interest has been devoted to the
study of linear gold atom chains (LAC)
\cite{Yan,Ohn,VRPRB,LandSc,Land,San,Kizu2}.
The invention and refinement of experimental techniques such as
scanning tunneling microscope (STM) \cite{LandSc,Pasc,Agra,Oles}, 
high resolution transmission
electron microscopy (HRTEM) \cite{Kizu,Kizu1,Kon1,Kon2,Ohn, VRPRB,VRPRL},
and mechanically 
controllable break
junction (MCBJ) \cite{Mull,Kra,VRPRL} 
have made possible the fabrication 
of such atomic size
contacts. As the LACs are in atomic scale, quantization of the conductance 
has been therefore observed (see recent review in \cite{Ruintbook}).

A surprising fact has been the observation of large bonds between gold atoms
in stable monoatomic gold chains with respect to the normal
nearest-neighbor distances in gold bulk (2.88 {\AA}), or in Au$_{2}$ 
dimers (2.5 {\AA}) \cite{Hak}. It has been reported \cite{Ohn,Yan,VRPRB,VRPRL,Takai} that interatomic distances
frequently vary in the 2.9 - 3.6 {\AA} range and also, extremely long distances close to 5 {\AA} 
have been rarely observed \cite{Kon2}. The origin of these large distances has
been a serious and unresolved challenge for theoretical interpretation.
Many theoretical studies have been reported using different techniques
\cite{Sor,Hak,Oka,San} such as molecular dynamics and density functional
theory (DFT). However, none of these works were able to reproduce the observed
large interatomic distances. The chain rupture is always observed for
distances significantly below the experimental values.

Neither of these theoretical works have considered the possibility of the
monoatomic chains being contaminated by some impurities \cite{Ohn,VRPRL,VRPRB,Kon2}
during sample fabrication. We have decided to explore this possibility,
in special considering carbon atom contamination that, from
experimental conditions, are most likely spurious atoms during HRTEM 
observations \cite{VRPRL,VRPRB}.
As recently demonstrated, the presence of carbon atoms would be
undetectable (because of the low contrast) in HRTEM images and
would appear as large unusual Au-Au distances \cite{Kon2}. 

In this paper we investigate both experimentally and theoretically the
formation of suspended gold chains, addressing the calculations
to analyze the origin of the large inter-gold distances experimentally observed.

We have generated stable linear gold atom chains
$in$ $situ$ in a HRTEM (JEM 3010 URP, 300 kV, 1.7 {\AA} resol., at LME/LNLS Campinas, Brazil) following the 
method developed by Kondo and Takanayagi \cite{Kon1}. This procedure is based on using electron-beam irradiation (current density of
100 A/cm$^{2}$) to drill holes at different sites of a self-supported gold thin
film (5 nm thick, average grain size 50-100 nm) until a nanometric neck
is formed between neighboring holes; the electron beam
intensity is then reduced ($\sim$30 A/cm$^{2}$) in order to perform the image
acquisition. This experimental approach has been carefully described elsewhere \cite{Rod3}. The presented images
were acquired from real time recording using a high sensitivity TV camera (Gatan 622SC) coupled to a 
conventional video recorder.

\begin{figure}
\includegraphics[width = 8 cm]{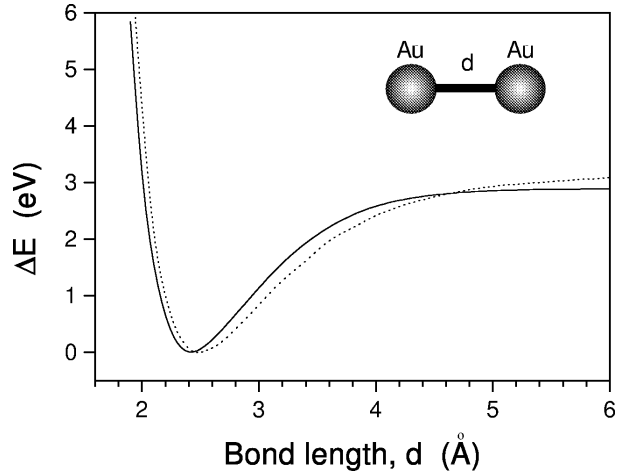}
\caption{Au$_2$ total energy (relative to the
minimum) as a function of the interatomic distance. Solid line (dot line) corresponds
to calculation with DMol (Titan) package.}
\end{figure}

In order to try to understand the origin of the unusual large Au-Au
interatomic distances molecular clusters were used to simulate LACs.
The geometry optimization calculations were
performed using $ab$ $initio$ density functional total energy methods.
We have used DMOL \cite{Dmol} and TITAN \cite{Titan} packages.
The density functional is treated by local density approximation with
a Wang-Perdew \cite{WP} (Vosko-Wilk-Nusair) \cite{VWN} 
exchange-correlation functional for DMOL (TITAN) case. For the
computation, we have used a double numerical (DND) basis set
together with polarization functions for DMOL, and a LACVP$^{**}$
basis set in the case of TITAN. The reliability and the method dependence of the
predicted geometries were tested carrying out a comparative study for
the dimeric interdistance. As we can see from Fig. 1, the qualitative
features are almost the same and the minimum predicted distance
(2.44 {\AA} for DMOL, and 2.49 {\AA} for TITAN calculations) are
in very good agreement with the experimental value of 2.47 {\AA} \cite{Hab,Yu}.

\begin{figure}
\includegraphics[width = 8 cm]{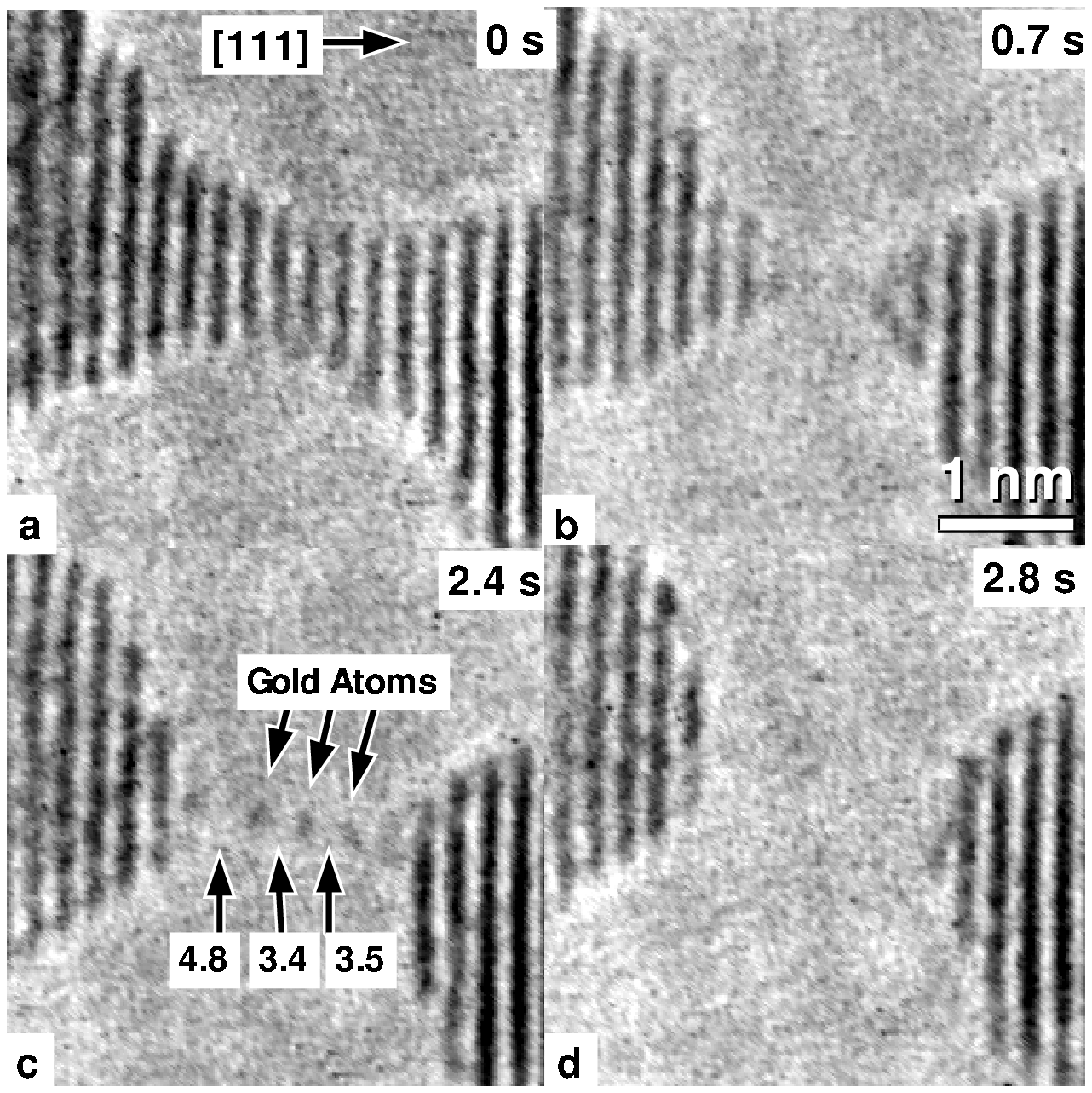}
\caption{(a)-(d) Nanowire time evolution when stretched approximately along the [111] direction. 
An LAC is formed in (c), where the Au-Au interatomic distances are indicated in {\AA} (see text for 
explanations). Atom positions appear dark.}
\end{figure}

In Fig. 2 we show a series of snapshots of the elongation of a gold nanowire. After forming a one-atom-thick 
constriction in Fig. 2b, further stretching induces the generation of an LAC consisting of three hanging atoms 
(Fig. 2(c)); finally, the constriction breaks and the apexes retract (Fig. 2(d)). This atom chain has been chosen 
because it displays a high contrast and also a peculiar interatomic distance distribution: a) two bonds within 
the frequently reported range (3.4 and 3.5 {\AA}, error bar 0.2 {\AA}) and; b) a extremely long bond 
attaining 4.8 {\AA}. Bonds in the 4.5-5 {\AA} interval have already been reported \cite{Kon1}, but although
they do occur, they have been observed only in very rare cases in our HRTEM experiments. 
It is interesting to observe the LAC interatomic distance histogram distribution (Fig. 3). 
Although the number of counts is rather low, we can easily observe that the bond-length values cover the whole range from 2.88 {\AA}
(nearest neighbor distance) to 3.7 {\AA}. The histogram shows a peak position at about 3.6 {\AA}, which 
seems to be a threshold value. In addition, an isolated point 
is measured at 4.8 {\AA}, corresponding to the image in Fig. 2(c).

The image in Fig. 2c, allows the clear identification of two quite different interatomic distances 
which coexist on the same LAC. Thus, they can not be related to the same phenomena, as for example 
tensile stress on clean gold atoms. This example suggests that the presence of contamination is one of 
the main issues to be analyzed, in particular considering that light atoms such as carbon would not be 
detectable within the signal-to-noise ratio of typical HRTEM images \cite{Kon2}. We have then proceeded to 
model the LAC structures considering the presence of carbon contamination. Since it has also been speculated 
that other atoms could be present \cite{Kon2}, we have analyzed N, O, S,
and Si atoms as well. However, carbon is the atom that provides results more consistent
when compared to the experimental data. 
Thus we present here only the C results.

\begin{figure}
\includegraphics[width = 8 cm]{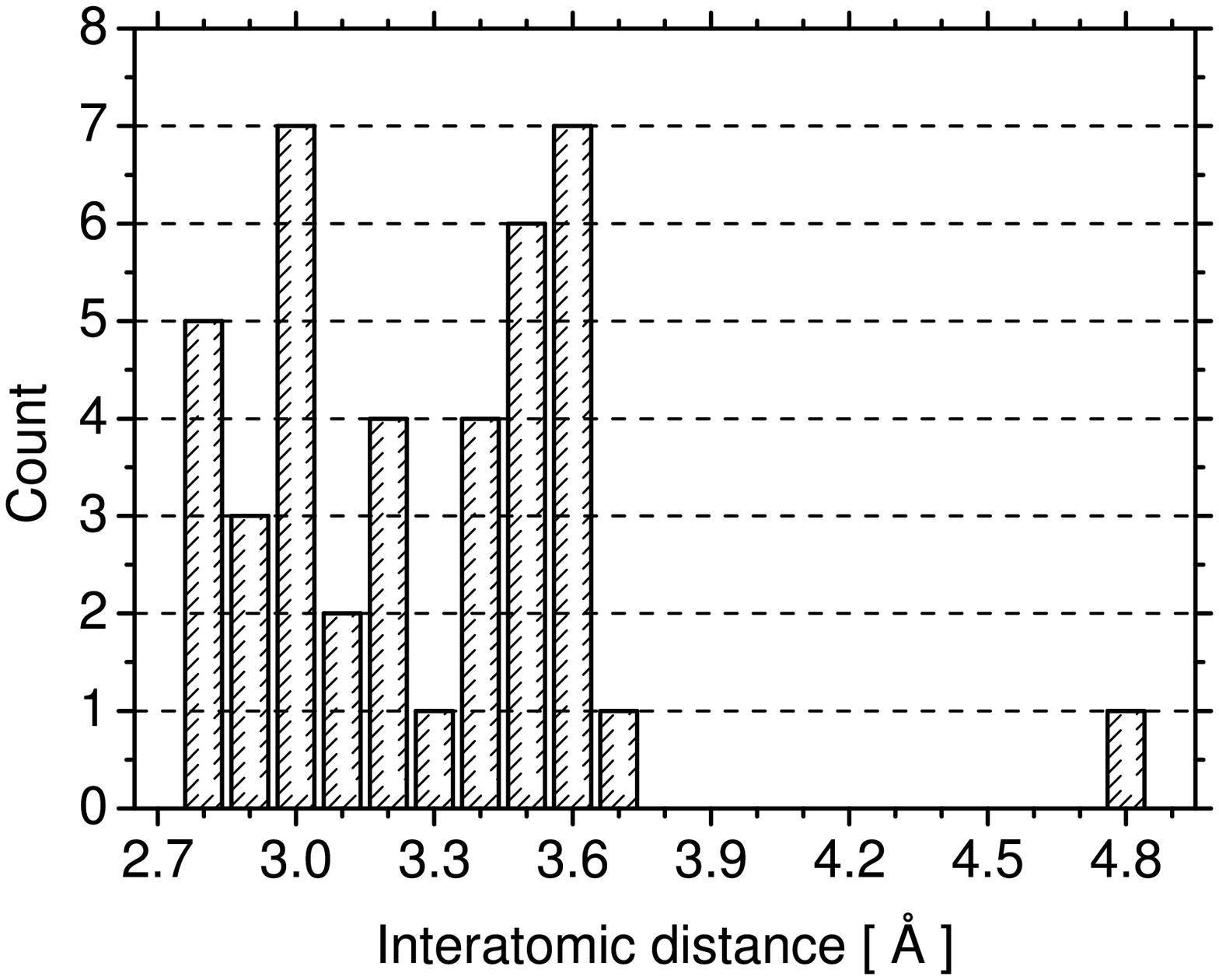}
\caption{Histogram on bond-lengths measured between hanging atom in gold LACs. A total of 41 
different distances are included. The analyzed LACs were generated from wires elongated along 
different crystallographic directions \cite{VRPRL}.}
\end{figure}

Initial structures were generated considering that the gold atoms are at arbitrary
positions forming linear chains at distances of $\sim$4.0 {\AA}.
The carbon atoms were positioned at distances of $\sim$1.5 {\AA}
above the middle chain positions. Geometrical optimizations were then carried
out. No constraints or symmetries
were imposed.
We have adopted the standard "naked" cluster approach, where no hydrogen or dummy
termination atoms are used \cite{Habe,Pac}. Recent ab-initio DFT calculations
for suspended gold chains \cite{San} showed that the geometries for 
"naked" finite and infinite (cyclic boundary conditions) structures are essentially the same.
This is an indication that the "naked" approach is appropriated for LACs. 

In Fig. 4, we present the results for some selected structures
obtained with DMOL; TITAN calculations produced similar results. 
The displayed examples are representative of possible structures 
considering the inclusion of carbon atoms at
different positions or configurations. We have considered C 
incorporation both in its atomic and 
dimeric forms. 
In the former case, C-Au and the Au-Au distances are of 1.8 and 3.6
{\AA}, respectively, and the structures are almost linear [Fig. 4(b)].
It is interesting to notice that the presence of a C atom between
gold atoms does not affect their next Au-Au distances [Fig. 4(c)]
which remains at the same value of the Au$_{2}$ [Fig. 4(a)].
The same behavior is observed for the C$_{2}$ incorporation, with the
exception that the Au-Au distances are now $\sim$5.0 {\AA}
[Figs. 4(d) and 4(e)]. To verify whether the presence of a large
number of terminal gold atoms would affect these bond-lengths, we have also
considered the case of adding two pyramidal (4 atoms each)
gold structures [Fig. 4(f)] at the chain termination. 
As we can see, the distance pattern is
almost the same of Fig. 4(e).

In Figs. 4(g) and 4(h), we show the obtained theoretical structures in
two situations, totally free to relax and with the length
geometrical constraint to match the total LAC length as observed in the experimental chain
reported in Fig. 3 of Ref. \cite{VRPRB}. For Fig. 4(g) the values are almost the same for similar
structures considered in Figs. 4(b) and 4(c). For the constrained
structure [Fig. 4(h)], we observe that the distances are slightly
affected (0.1 {\AA} reduction) and can be attributed to the
observed bending. Considering the experimental error bar (0.2 {\AA}),
the agreement between the experimental and theoretical data is
excellent, providing strong support to the presence of C atoms as
proposed. In Fig. 4(i) we show a predicted structure for the
case of an unstrained linear chain of four gold atoms including dimeric and atomic C incorporation. 
This system is to be compared with the LAC of Fig. 2(c). Again, considering the error bar there 
is an excellent match between the experimental and theoretical data.

\begin{figure}
\includegraphics[width = 7.5 cm]{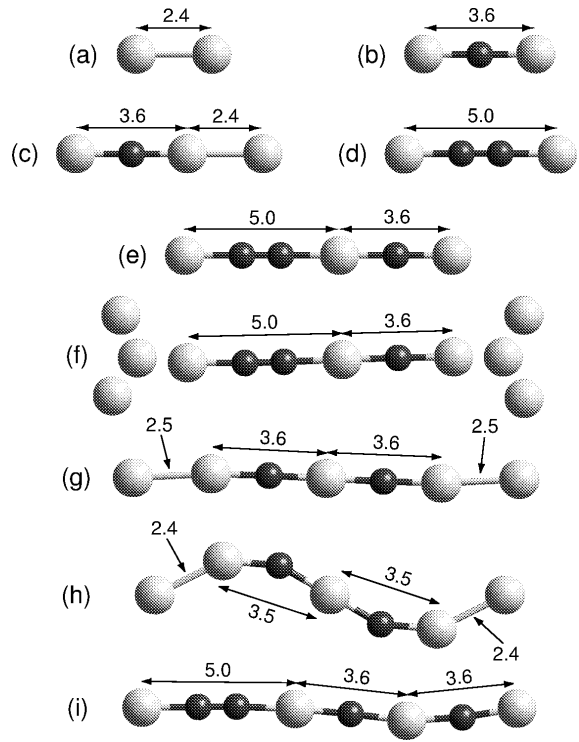}
\caption{Optimized gold chains with carbon impurities, for selected configurations.
The chain (a) is a gold dimer. The other chains, except (f), are bidimensional
molecules, contained in the figure plane. The system in (f) is a gold-carbon
chain, with suspended tips in the extremes, composed by four gold atoms in a
near (100)-Au geometry. Light (dark grey) spheres correspond to gold
(carbon) atoms in the chain. Distances are in {\AA}.} 
\end{figure}

The theoretical calculations presented above have mainly addressed the effect of carbon 
incorporation into the suspended atom chains. We would like to stress 
that the existence of distances in the range of 3.0-3.6 {\AA}
do not necessarily demand the existence of C impurities. This can be deduced from the bond-length histogram 
in Fig. 3 which shows (within available low statistics) a continuous range from the expected gold 
distance in the bulk form to 3.7 {\AA}, as it should be expected for tensile stress elongation applied on 
clean bonds. However the histogram peak at about 3.5 {\AA} could result as a direct effect of the C 
presence. An alternative explanation for the more frequent observation of 3.5 {\AA} interatomic 
distance can be that this value is the threshold for elongation \cite{Pyy} and, further increase would require 
larger tensile stress. At this point, the apexes start to absorb the elastic energy because they 
have larger degrees of freedom, as predicted in molecular dynamic simulations \cite{Zach,RubioB}. 
Thus, the more frequent occurrence of the 3.6 {\AA} bonds is due to the fact that this length 
has a longer life time and then allow better imaging conditions during the experiments. 
The experimental data do not allows us to rule out the two situations (clean or contaminated).
However, the histogram profile seems to be more compatible with the coexistence of the two regimes.

As for the longest Au-Au distances of $\sim$4.0-5.0 {\AA} our
results strongly indicate that they should not exist (as evidenced
by many theoretical works) \cite{Sor,Oka,San, Hak}. To explain
these distances we must invoke the presence of
impurity contamination, more specifically C$_{2}$. 
There are many possible sources of contamination in transmission electron
microscopy work, such as vacuum oil from rotatory
and diffusion pumps, grease and rubber O-rings, etc..
Concerning the physical origin, radiation damage of adsorbed
hydrocarbon molecules causes a carbon rich,
polymerized film to form and grow on the electron irradiated areas 
(for details, see for instance Ref. \cite{Trans}).
In our experiments the initial carbon layer is cleaned by the intense
irradiation (see Refs. \cite{VRPRB,VRPRL,Rod3}, however
during the nanowire rupture the
vacuum chamber contains hydrocarbon molecules from the residual
gas pressure (10-7 torr in our HRTEM). Although these molecules are in low
density, they slowly and continuously deposit carbon atoms on the nanowires
surface. When these C atoms are generated close to the junction region (where
the e-beam is focused) they
may be incorporated into the suspended atom chains.
The remarkable agreement between experimental and our theoretical
data, associated with the fact that the presence of C atoms during the
experiment is highly probable and should not be detectable, support the structures we are proposing.

In summary, we have experimentally observed the formation of stable
linear gold atom chains, and get direct real-space information on
atomic positions and bond-lengths. To explain the origin of large
gold-gold interdistances in those linear chains, we have 
carried out $ab$ $initio$ density functional theory
geometrical optimization calculations for gold clusters assuming
the presence of C atoms as impurity contamination. Our results
show that the apparent puzzle of the existence of long 
interatomic Au-Au distances (4.0-5.0 {\AA}) can be easily explained by
the presence of C$_{2}$ impurities. However, the bond-lengths above 3.0 {\AA} might be due to a mixture of clean 
stressed bonds and those contaminated by incorporating a single carbon atom. In principle these same arguments are
applicable to other metallic monoatomic nanowires (formed under
similar experimental conditions) such as Pt \cite{VRPt} and Ag \cite{VRAg}, whose
structures should also present large intermetallic distances.


Work supported in part by the Brazilian agencies CNPq, FINEP, and FAPESP.
The authors also wish to acknowledge support from Accelrys, Inc. for
helpful assistance.

\end{document}